\documentstyle[twoside,psfig]{article}

\catcode`\@=11
\long\def\@makefntext#1{
\protect\noindent \hbox to 3.2pt {\hskip-.9pt  
$^{{\eightrm\@thefnmark}}$\hfil}#1\hfill}               

\def\thefootnote{\fnsymbol{footnote}}
\def\@makefnmark{\hbox to 0pt{$^{\@thefnmark}$\hss}}    
        
\def\ps@myheadings{\let\@mkboth\@gobbletwo
\def\@oddhead{\hbox{}
\rightmark\hfil\eightrm\thepage}   
\def\@oddfoot{}\def\@evenhead{\eightrm\thepage\hfil
\leftmark\hbox{}}\def\@evenfoot{}
\def\sectionmark##1{}\def\subsectionmark##1{}}

\oddsidemargin=\evensidemargin
\renewcommand{\thefootnote}{\fnsymbol{footnote}}

\newcounter{sectionc}\newcounter{subsectionc}\newcounter{subsubsectionc}
\renewcommand{\section}[1] {\vspace{12pt}\addtocounter{sectionc}{1} 
\setcounter{subsectionc}{0}\setcounter{subsubsectionc}{0}\noindent 
        {\tenbf\thesectionc. #1}\par\vspace{5pt}}
\renewcommand{\subsection}[1] {\vspace{12pt}\addtocounter{subsectionc}{1} 
        \setcounter{subsubsectionc}{0}\noindent 
        {\bf\thesectionc.\thesubsectionc. {\kern1pt \bfit #1}}\par\vspace{5pt}}
\renewcommand{\subsubsection}[1] {\vspace{12pt}\addtocounter{subsubsectionc}{1}
        \noindent{\tenrm\thesectionc.\thesubsectionc.\thesubsubsectionc.
        {\kern1pt \tenit #1}}\par\vspace{5pt}}
\newcommand{\nonumsection}[1] {\vspace{12pt}\noindent{\tenbf #1}
        \par\vspace{5pt}}

\newcounter{appendixc}
\newcounter{subappendixc}[appendixc]
\newcounter{subsubappendixc}[subappendixc]
\renewcommand{\thesubappendixc}{\Alph{appendixc}.\arabic{subappendixc}}
\renewcommand{\thesubsubappendixc}
        {\Alph{appendixc}.\arabic{subappendixc}.\arabic{subsubappendixc}}

\renewcommand{\appendix}[1] {\vspace{12pt}
        \refstepcounter{appendixc}
        \setcounter{figure}{0}
        \setcounter{table}{0}
        \setcounter{lemma}{0}
        \setcounter{theorem}{0}
        \setcounter{corollary}{0}
        \setcounter{definition}{0}
        \setcounter{equation}{0}
        \renewcommand{\thefigure}{\Alph{appendixc}.\arabic{figure}}
        \renewcommand{\thetable}{\Alph{appendixc}.\arabic{table}}
        \renewcommand{\theappendixc}{\Alph{appendixc}}
        \renewcommand{\thelemma}{\Alph{appendixc}.\arabic{lemma}}
        \renewcommand{\thetheorem}{\Alph{appendixc}.\arabic{theorem}}
        \renewcommand{\thedefinition}{\Alph{appendixc}.\arabic{definition}}
        \renewcommand{\thecorollary}{\Alph{appendixc}.\arabic{corollary}}
        \renewcommand{\theequation}{\Alph{appendixc}.\arabic{equation}}
        \noindent{\tenbf Appendix \theappendixc #1}\par\vspace{5pt}}
\newcommand{\subappendix}[1] {\vspace{12pt}
        \refstepcounter{subappendixc}
        \noindent{\bf Appendix \thesubappendixc. {\kern1pt \bfit #1}}
        \par\vspace{5pt}}
\newcommand{\subsubappendix}[1] {\vspace{12pt}
        \refstepcounter{subsubappendixc}
        \noindent{\rm Appendix \thesubsubappendixc. {\kern1pt \tenit #1}}
        \par\vspace{5pt}}

\newcommand{\textlineskip}{\baselineskip=13pt}
\newcommand{\smalllineskip}{\baselineskip=10pt}

\def\eightcirc{
\begin{picture}(0,0)
\put(4.4,1.8){\circle{6.5}}
\end{picture}}
\def\eightcopyright{\eightcirc\kern2.7pt\hbox{\eightrm c}} 

\newcommand{\copyrightheading}[1]
        {\vspace*{-2.5cm}\smalllineskip{\flushleft
        {\footnotesize Modern Physics Letters A, #1}\\
        {\footnotesize $\eightcopyright$\, World Scientific Publishing
         Company}\\
         }}

\newcommand{\publisher}[2]{{\begin{center}\footnotesize\smalllineskip 
        Received #1\\
        Revised #2
        \end{center}
        }}

\def\abstracts#1#2#3{{
        \centering{\begin{minipage}{4.5in}\footnotesize\baselineskip=10pt
        \parindent=0pt #1\par 
        \parindent=15pt #2\par
        \parindent=15pt #3
        \end{minipage}}\par}} 

\renewenvironment{thebibliography}[1]
        {\frenchspacing
         \ninerm\baselineskip=11pt
         \begin{list}{\arabic{enumi}.}
        {\usecounter{enumi}\setlength{\parsep}{0pt}     
         \setlength{\leftmargin 12.7pt}{\rightmargin 0pt} 
         \setlength{\itemsep}{0pt} \settowidth
        {\labelwidth}{#1.}\sloppy}}{\end{list}}

\newcounter{itemlistc}
\newcounter{romanlistc}
\newcounter{alphlistc}
\newcounter{arabiclistc}

\newcommand{\fcaption}[1]{
        \refstepcounter{figure}
        \setbox\@tempboxa = \hbox{\footnotesize Fig.~\thefigure. #1}
        \ifdim \wd\@tempboxa > 5in
           {\begin{center}
        \parbox{5in}{\footnotesize\smalllineskip Fig.~\thefigure. #1}
            \end{center}}
        \else
             {\begin{center}
             {\footnotesize Fig.~\thefigure. #1}
              \end{center}}
        \fi}

\newcommand{\tcaption}[1]{
        \refstepcounter{table}
        \setbox\@tempboxa = \hbox{\footnotesize Table~\thetable. #1}
        \ifdim \wd\@tempboxa > 5in
           {\begin{center}
        \parbox{5in}{\footnotesize\smalllineskip Table~\thetable. #1}
            \end{center}}
        \else
             {\begin{center}
             {\footnotesize Table~\thetable. #1}
              \end{center}}
        \fi}

\def\@citex[#1]#2{\if@filesw\immediate\write\@auxout
        {\string\citation{#2}}\fi
\def\@citea{}\@cite{\@for\@citeb:=#2\do
        {\@citea\def\@citea{,}\@ifundefined
        {b@\@citeb}{{\bf ?}\@warning
        {Citation `\@citeb' on page \thepage \space undefined}}
        {\csname b@\@citeb\endcsname}}}{#1}}

\newif\if@cghi
\def\cite{\@cghitrue\@ifnextchar [{\@tempswatrue
        \@citex}{\@tempswafalse\@citex[]}}
\def\citelow{\@cghifalse\@ifnextchar [{\@tempswatrue
        \@citex}{\@tempswafalse\@citex[]}}
\def\@cite#1#2{{$\null^{#1}$\if@tempswa\typeout
        {IJCGA warning: optional citation argument 
        ignored: `#2'} \fi}}

\def\pmb#1{\setbox0=\hbox{#1}
        \kern-.025em\copy0\kern-\wd0
        \kern.05em\copy0\kern-\wd0
        \kern-.025em\raise.0433em\box0}

\def\fnt#1#2{\footnotetext{\kern-.3em
        {$^{\mbox{\scriptsize #1}}$}{#2}}}

\def\fpage#1{\begingroup
\voffset=.3in
\thispagestyle{empty}\begin{table}[b]\centerline{\footnotesize #1}
        \end{table}\endgroup}

\def\runninghead#1#2{\pagestyle{myheadings}
\markboth{{\protect\footnotesize\it{\quad #1}}\hfill}
{\hfill{\protect\footnotesize\it{#2\quad}}}}
\font\tenrm=cmr10
\font\tenit=cmti10 
\font\tenbf=cmbx10
\font\bfit=cmbxti10 at 10pt
\font\ninerm=cmr9

\font\eightrm=cmr8

\def\qed{\hbox{${\vcenter{\vbox{                        
   \hrule height 0.4pt\hbox{\vrule width 0.4pt height 6pt
   \kern5pt\vrule width 0.4pt}\hrule height 0.4pt}}}$}}

\renewcommand{\thefootnote}{\fnsymbol{footnote}}        

\begin{document}
\setlength{\textheight}{7.7truein}  

\runninghead{L.~Sriramkumar}{Odd Statistics in Odd Dimensions 
for Odd Couplings}

\normalsize\textlineskip
\thispagestyle{empty}
\setcounter{page}{1}

\copyrightheading{}                     

\vspace*{0.88truein}

\fpage{1}
\centerline{\bf ODD STATISTICS IN ODD DIMENSIONS}
\baselineskip=13pt
\centerline{\bf FOR ODD COUPLINGS\footnote{This title is motivated by the title of 
an unpublished paper by C.~R.~Stephens\cite{stephens85}.}}
\vspace*{0.37truein}
\centerline{\footnotesize L.~SRIRAMKUMAR\footnote{Present 
address:~Harish-Chandra Research Institute, Chhatnag Road, 
Jhunsi, Allahabad 211 019, India. E-mail:~sriram@mri.ernet.in.}}
\baselineskip=12pt
\centerline{\footnotesize\it Theoretical Physics Institute,
Department of Physics, University of Alberta}
\baselineskip=10pt
\centerline{\footnotesize\it Edmonton, Alberta~T6G~2J1, 
Canada.}
\vspace*{10pt}

\publisher{(received date)}{(revised date)}

\vspace*{0.21truein}
\abstracts{We consider the response of a uniformly accelerated 
monopole detector that is coupled {\it non-linearly}\/ to 
the~$n$th power of a quantum scalar field in $(D+1)$\/-dimensional 
flat spacetime. 
We show that, when $(D+1)$\/ is even, the response of the detector 
in the Minkowski vacuum is characterized by a Bose-Einstein factor
for all~$n$.\/ 
Whereas, when $(D+1)$\/ is odd, we find that a Fermi-Dirac factor 
appears in the detector response when~$n$\/ is odd, but a Bose-Einstein 
factor arises when~$n$\/ is even.
We emphasize the point that, since, along the accelerated
trajectory, the Wightman function and, as a result, the 
$(2n)$\/-point function satisfy the Kubo-Martin-Schwinger 
condition (as required for a scalar field) in {\it all}\/ 
dimensions, the appearance of a Fermi-Dirac factor (instead 
of the expected Bose-Einstein distribution) for odd~$(D+1)$\/ 
{\it and}\/~$n$\/ reflects a peculiar feature of the detector rather 
than imply a fundamental change in field theory.}{}{}



\vspace*{1pt}\textlineskip      
\section{Introduction}          
\vspace*{-0.5pt}
\noindent
It has been a quarter of a century now since it was discovered 
that the response of a uniformly accelerated monopole 
detector that is coupled to a quantized massless scalar field 
is characterized by a Planckian distribution when the field is 
assumed to be in the Minkowski vacuum\cite{unruh76,dewitt79}.
However, about a decade after the original discovery, it was  
noticed that this result is true only in even-dimensional flat 
spacetimes and it was pointed out that a Fermi-Dirac factor 
(rather than a Bose-Einstein factor) appears in the response of 
the accelerated detector when the dimensionality of spacetime 
is odd (see 
Refs.\cite{takagi84,takagi85a,stephens86,ooguri86,unruh86,takagi86}; 
for relatively recent discussions, see 
Refs.\cite{anglin93,terashima99}). 
The detector due to Unruh\cite{unruh76} and DeWitt\cite{dewitt79}
is coupled {\it linearly}\/ to the quantum scalar field.
During the last decade or so, motivated by different reasons,
there has been an occasional interest in literature in studying 
the response of detectors that are coupled {\it non-linearly}\/ 
to the quantum 
field\cite{hinton83,hinton84,paddytp87,suzuki97,sriram99}.
It will be interesting to examine whether the non-linearity 
of the coupling affects the result in odd-dimensional flat 
spacetimes that we mentioned above.

In this note, we shall consider the response of a uniformly 
accelerated monopole detector that is coupled to the $n$th 
power (with $n$ being a positive integer) of a quantum scalar 
field in $(D+1)$-dimensional flat spacetime.
As we shall see, the non-linearity of the detector's coupling 
affects the afore-mentioned result in odd spacetime dimensions 
in an interesting fashion.
We shall show that, when $(D+1)$ is even, a Bose-Einstein 
factor arises in the response of the detector for all $n$, 
whereas, when $(D+1)$ is odd, a Fermi-Dirac factor appears 
in the detector response when~$n$ is odd, but a Bose-Einstein 
factor arises when $n$ is even.

We shall adopt units such that $\hbar=c=1$ and we shall denote 
the set of $(D+1)$ coordinates $x^{\mu}$ as ${\tilde x}$.

\setcounter{footnote}{0}
\renewcommand{\thefootnote}{\alph{footnote}}

\section{The Non-linearly Coupled Detector}
\noindent
Consider a monopole detector that is moving along a trajectory 
${\tilde x}(\tau)$, where $\tau$ is the proper time in the frame 
of the detector.
Let the detector interact with a real scalar field~$\Phi$ through 
the non-linear interaction Lagrangian\cite{suzuki97}
\begin{equation}
{\cal L}_{\rm int}
= {\bar c}\, m(\tau)\; \Phi^n\left[{\tilde x}(\tau)\right],
\label{eqn:nlint}
\end{equation}
where ${\bar c}$ is a small coupling constant, $m(\tau)$ 
is the detector's monopole moment and~$n$ is a positive
integer that denotes the index of non-linearity of 
the coupling.
Let us now assume that the quantum field ${\hat \Phi}$ is 
initially in the vacuum state $\left\vert 0 \right\rangle$ 
and the detector is in its ground state~$\left\vert E_0 
\right\rangle$ 
corresponding to an energy eigen value~$E_0$.
Then, up to the first order in perturbation theory, the 
amplitude of transition of the non-linearly coupled detector 
to an excited state~$\left\vert E \right\rangle$, corresponding 
to an energy eigen value~$E\, \left(>E_0\right)$, is described 
by the integral (see, for e.g., Ref.\cite{bd82})
\begin{equation}
{\cal A}_{n}({\cal E}) = {\cal M}
\int\limits_{-\infty}^{\infty} d\tau\, e^{i {\cal E}\tau}\, 
\left\langle\Psi\right\vert\, {\hat \Phi}^n[{\tilde x}(\tau)]\,
\left\vert 0\right\rangle,
\label{eqn:detamp}
\end{equation}
where ${\cal M}\equiv \left(i{\bar c}\, \left\langle E \right\vert 
{\hat m}(0)\left\vert E_{0} \right\rangle\right)$, ${\cal E}
=\left(E-E_0\right)>0$ and $\left\vert \Psi \right\rangle$ is 
the state of the quantum scalar field after its interaction 
with the detector.
(Since the quantity ${\cal M}$ depends only on the internal 
structure of the detector and does not depend on its motion, 
we shall drop this quantity hereafter.) 

The transition amplitude~${\cal A}_{n}({\cal E})$ above 
involves products of the quantum field~${\hat \Phi}$ at the 
{\it same}\/ spacetime point and, hence, we will encounter 
divergences when evaluating this transition amplitude.
In order to avoid the divergences, we shall normal order 
the operators in the matrix element in the transition 
amplitude~${\cal A}_{n}({\cal E})$ with respect to the 
Minkowski vacuum\cite{suzuki97}.
That is, we shall assume that the transition 
amplitude~(\ref{eqn:detamp}) above is instead 
given by the expression
\begin{equation}
{\bar {\cal A}}_{n}({\cal E}) 
=\int\limits_{-\infty}^{\infty} d\tau\, e^{i {\cal E}\tau}\, 
\left\langle\Psi\right\vert
:{\hat \Phi}^n[{\tilde x}(\tau)]:\left\vert 0\right\rangle,
\label{eqn:nodetamp}
\end{equation}
where the colons denote normal ordering with respect to the
Minkowski vacuum.
Then, the transition probability of the detector to all 
possible final states $\left\vert \Psi\right\rangle$ of 
the quantum field is given by
\begin{equation}
{\cal P}_{n}({\cal E}) 
=\sum_{\vert\Psi\rangle}{\vert 
{\bar {\cal A}}_{n}({\cal E})\vert}^2
= \int\limits_{-\infty}^\infty d\tau\, 
\int\limits_{-\infty}^\infty d\tau'\, 
e^{-i{\cal E}(\tau-\tau')}\, 
G^{(n)}\left[{\tilde x}(\tau), {\tilde x}(\tau')\right],
\label{eqn:detprob}
\end{equation}
where $G^{(n)}\left[{\tilde x}(\tau), {\tilde x}(\tau')\right]$ 
is the $(2n)$-point function defined as
\begin{equation}
G^{(n)}\left[{\tilde x}(\tau), {\tilde x}(\tau')\right]
=\left\langle 0 \right\vert 
:{\hat \Phi}^n\left[{\tilde x}(\tau)\right]:\,
:{\hat \Phi}^n\left[{\tilde x}(\tau')\right]:
\left\vert 0 \right\rangle.\label{eqn:2nptfn}
\end{equation}
In cases wherein the  $(2n)$-point function 
$G^{(n)}\left(\tau, \tau'\right) \left(\equiv G^{(n)}
\left[{\tilde x}(\tau), {\tilde x}(\tau')\right]\right)$ is 
invariant under time translations in frame of the detector, 
we can define a transition probability rate for the detector 
as follows:
\begin{equation}
{\cal R}_{n}({\cal E}) 
= \int\limits_{-\infty}^\infty d{\bar \tau}\;
e^{-i{\cal E}{\bar \tau}}\; G^{(n)}({\bar \tau}),
\label{eqn:detrate}
\end{equation} 
where ${\bar \tau}=(\tau -\tau')$.

\section{Odd Statistics in Odd dimensions for Odd Couplings}
\noindent
Let us now assume that the quantum scalar field~${\hat \Phi}$ 
is in the Minkowski vacuum.
In such a case, the $(2n)$-point function $G^{(n)}\left({\tilde x}, 
{\tilde x'}\right)$ simplifies to
\begin{equation}
G^{(n)}_{\rm M}\left({\tilde x}, {\tilde x'}\right)
= \left(n!\right)\, \left[G^{+}_{\rm M}\left({\tilde x}, 
{\tilde x'}\right)\right]^n,\label{eqn:2nptfnmv}
\end{equation}
where $G^{+}_{\rm M}\left({\tilde x}, {\tilde x'}\right)$ 
denotes the Wightman function in the Minkowski vacuum\footnote{It 
ought to be noted here that we would have arrived at the 
expression~(\ref{eqn:2nptfnmv}) for the $(2n)$-point function in 
the Minkowski vacuum even if we had started with the transition 
amplitude~(\ref{eqn:detamp}) (instead of the normal ordered 
amplitude~(\ref{eqn:nodetamp})), rewritten the resulting $(2n)$-point 
function in the transition probability in terms of the two-point 
functions using Wick's theorem and then replaced the divergent 
terms that arise (i.e. those two-point functions with coincident 
points) with the corresponding regularized expressions (for a 
discussion on this point, also see Ref.\cite{suzuki97}).}.
In $(D+1)$ spacetime dimensions (and for $(D+1)\ge 3$), the 
Wightman function for a massless scalar field in the Minkowski 
vacuum is given by (see, for instance, 
Refs.\cite{takagi85a,unruh86,takagi86})
\begin{equation}
G^{+}_{\rm M}({\tilde x},{\tilde x'})
= {\cal C}_{D}\; \left[(-1)\;
\left((t-t'-i\epsilon)^2
-\vert {\bf x}-{\bf x}'\vert^2\right)\right]^{-(D-1)/2},
\label{eqn:wgfnmvine}
\end{equation}
where $\epsilon\to 0^{+}$, $\left[t, {\bf x}\equiv\left(x^{1},
x^{2},\ldots,x^{D}\right)\right]$ are the Minkowski coordinates 
and the quantity~${\cal C}_{D}$ is given by  
\begin{equation}
{\cal C}_{D}= \left(4\pi^{(D+1)/2}\right)^{-1}\;
\Gamma\left[(D-1)/2\right]
\label{eqn:cd}
\end{equation}
with $\Gamma\left[(D-1)/2\right]$ denoting the Gamma function.

Now, consider a detector accelerating uniformly along the 
$x^{1}$~direction with a proper acceleration~$g$. 
The trajectory of such a detector is given by (see, for e.g.,
Ref.\cite{bd82})
\begin{equation}
t(\tau)=g^{-1}\, {\rm sinh}(g\tau)\;\;,\;\; 
x^{1}(\tau)=g^{-1}\, {\rm cosh}(g\tau)\;\;,\;\;
x^{2}=x^{3}=\ldots=x^{D}=0,
\label{eqn:acctraj}
\end{equation}
where~$\tau$ is the proper time in the frame of the detector.
On substituting this trajectory in the Minkowski Wightman 
function~(\ref{eqn:wgfnmvine}), we obtain that (see, for 
instance, Refs.\cite{takagi85a,unruh86,takagi86})
\begin{equation}
G_{\rm M}^{+}\left({\bar \tau}\right)
=\left[{\cal C}_{D}\; (g/2i)^{(D-1)}\right]\;
\biggl({\rm sinh}\left[\left(g{\bar \tau}/2\right)
-i\epsilon\right]\biggl)^{-(D-1)}.
\label{eqn:wgfnmvrind}
\end{equation}
Therefore, along the trajectory of the uniformly 
accelerated detector, the $(2n)$-point function in 
the Minkowski vacuum~(\ref{eqn:2nptfnmv}) is given by
\begin{equation}
G^{(n)}_{\rm M}\left({\bar \tau}\right)
=(n!)\;\left[{\cal C}_{D}^{n}\; (g/2i)^{\alpha}\right]\; 
\biggl({\rm sinh}\left[\left(g{\bar \tau}/2\right)
-i\epsilon\right]\biggl)^{-\alpha},
\label{eqn:2nptfnmvrind}
\end{equation}
where $\alpha=\left[(D-1)n\right]$.

On substituting the $(2n)$-point 
function~(\ref{eqn:2nptfnmvrind}) in the 
expression~(\ref{eqn:detrate}) and carrying 
out the resulting integral, we find that the transition 
probability rate of the uniformly accelerated, non-linearly
coupled detector can be written as (cf.~Ref.\cite{gr80}, p.~305, 
Eq.~3.314; p.~950, Eq.~8.384.1; p.~937, Eqs.~8.331, 8.332.1 
and 8.332.2)
\begin{equation}
{\cal R}_{n}({\cal E}) 
={\cal B}(n,D)\;\;
\left\{
\begin{array}{l}
\left(g^{\alpha}/{\cal E}\right)\;
{\underbrace{\left[\exp(2\pi{\cal E}/g)-1\right]^{-1}}}\;
\prod\limits_{l=0}^{(\alpha-2)/2}
\left[l^2+({\cal E}/g)^2\right]\\
\qquad\quad\,\mbox{Bose-Einstein factor}\qquad
\qquad\qquad\quad\;\mbox{when $\alpha$ is even}\\
g^{(\alpha-1)}\;\;
{\underbrace{\left[\exp(2\pi{\cal E}/g)+1\right]^{-1}}}\;
\prod\limits_{l=0}^{(\alpha-3)/2}
\left[((2l+1)/2)^2+({\cal E}/g)^2\right]\\
\qquad\quad\;\mbox{Fermi-Dirac factor}\qquad
\qquad\qquad\qquad\mbox{when $\alpha$ is odd,}
\end{array}\right.
\label{eqn:fnlexp}
\end{equation}
where the quantity ${\cal B}(n,D)$ is given by
\begin{equation}
{\cal B}(n,D)=(2\pi)\,(n!)\; 
\left[{\cal C}_{D}^{n}/\Gamma(\alpha)\right].
\end{equation}
When $(D+1)$ is even, $\alpha$ is even for all $n$ and, hence, 
a Bose-Einstein factor will always arise in the response of 
the uniformly accelerated detector in an even-dimensional flat 
spacetime.
Whereas, when $(D+1)$ is odd, evidently, $\alpha$ will be odd 
or even depending on whether $n$ is odd or even.
Therefore, in an odd-dimensional flat spacetime, a Fermi-Dirac 
factor will arise in the detector response when~$n$ is odd (as
in the case of the Unruh-DeWitt detector), but a Bose-Einstein 
factor will appear when~$n$ is even!

\begin{figure}[htbp] 
\vspace*{13pt}
$\qquad{\bar {\cal R}}_{n}({\cal E})$
\centerline{\psfig{file=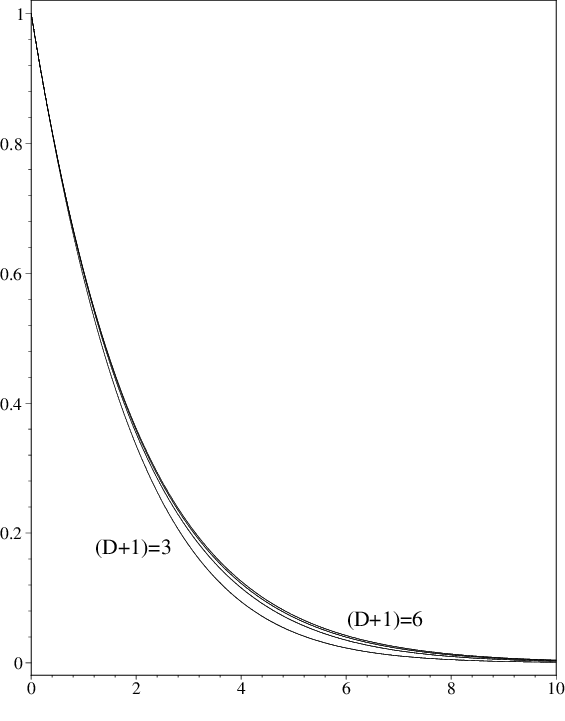}} 
\hskip -50 pt ${\cal E}$
\vspace*{13pt}
\fcaption{${\bar {\cal R}}_{n}({\cal E})\,\, 
{\rm vs.}\,\,{\cal E}$ for $n=3$ and $(D+1)=3,4,5,6$.}\label{fig:n3}
\end{figure}

\begin{figure}[htbp] 
\vspace*{13pt}
$\qquad{\bar {\cal R}}_{n}({\cal E})$
\centerline{\psfig{file=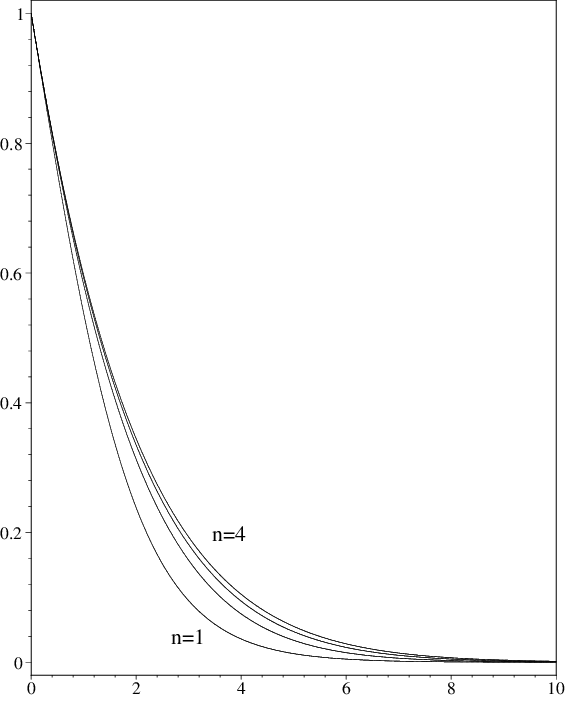}} 
\hskip -50 pt ${\cal E}$
\vspace*{13pt}
\fcaption{${\bar {\cal R}}_{n}({\cal E})\,\, 
{\rm vs.}\,\, {\cal E}$ for $(D+1)=3$ and $n=1,2,3,4$.}\label{fig:D2}
\end{figure}

A few remarks regarding this curious result are in order.
Firstly, the temperature associated with the Bose-Einstein and 
the Fermi-Dirac factors that appear in the response of the
non-linearly coupled detector is the standard Unruh temperature, 
viz.~$(g/2\pi)$. 
Secondly, the response of the detector is characterized 
{\it completely}\/ by either a Bose-Einstein or a Fermi-Dirac 
distribution {\it only}\/ in situations wherein $\alpha<3$.
For cases such that $\alpha\ge 3$, the detector response contains,
in addition to a Bose-Einstein or a Fermi-Dirac factor, a term 
which is polynomial in $({\cal E}/g)$.
Thirdly, in Figs.~\ref{fig:n3} and~\ref{fig:D2}, 
following Unruh\cite{unruh86}, we have plotted the transition 
probability rate of the detector (in fact, the quantity 
${\bar {\cal R}}_{n}({\cal E})=\left[{\cal R}_{n}({\cal E})/
{\cal R}_{n}(0)\right]$) for a few different values of~$D$ for 
the case $n=3$ and for a few different values of~$n$ for the 
case wherein $(D+1)=3$, respectively.
(In plotting these figures, we have set $g=(2\pi)$.)
It is interesting to note from these two figures that, though the 
characteristic response of the detector alternates between the 
Bose-Einstein and the Fermi-Dirac factors as we go from one $D$ 
to another for odd $n$ (or from one $n$ to another when $(D+1)$ 
is odd), the complete spectra themselves exhibit a smooth  
dependence both on the index of non-linearity of the coupling and 
the dimension of spacetime.

\section{{\it Apparent}\/ nature of the odd statistics}
\noindent
The point that needs to be emphasized regarding the result 
we have obtained above is the {\it apparent}\/ nature of the 
``inversion of statistics'' that arises in odd dimensions 
for odd couplings.
It is easy to see that, in the frame of the uniformly 
accelerated detector, the Wightman function in the Minkowski 
vacuum~(\ref{eqn:wgfnmvrind}) is skew-periodic in imaginary 
proper time with a period corresponding to the inverse of 
the Unruh temperature, i.e.
\begin{equation} 
G_{\rm M}^{+}\left({\bar \tau}\right)
=G_{\rm M}^{+}\left[-{\bar \tau}+i(2\pi/g)\right].
\end{equation}
Also, this property holds for all $D$.
In other words, along the accelerated trajectory, the Wightman 
function satisfies the Kubo-Martin-Schwinger (KMS) condition as 
required for a Bosonic field in {\it all}\/ dimensions (for a 
discussion on the KMS condition, see, for e.g., 
Ref.\cite{takagi86}, Chaps.~4 and~5).
Since the $(2n)$-point function in the Minkowski 
vacuum is proportional to the $n$th power of the 
Wightman function, it is then obvious that, in the 
frame of the accelerated detector, the $(2n)$-point 
function will also be skew-periodic in imaginary 
proper time, thereby satisfying the KMS condition 
(as required for a Bosonic field) for {\it all}\/ $D$ 
and $n$ (cf. Eq.~(\ref{eqn:2nptfnmvrind})). 
Had the appearance of the Fermi-Dirac factor for odd $(D+1)$
and $n$ been the manifestation of a truly fundamental change 
in the field theory, then, in such situations, the $(2n)$-point 
function along the accelerated trajectory would have been 
skew {\it as well as}\/ anti-periodic in imaginary proper time 
as expected of a Fermionic field (see, for instance, 
Ref.\cite{ooguri86}).
The fact that this does not occur then clearly implies that the 
appearance of the Fermi-Dirac factor (instead of the expected 
Bose-Einstein factor) for odd $(D+1)$ and $n$ just reflects a 
peculiar feature of the detector rather than suggest a 
fundamental shift in the field theory.

\section{Outlook}
\noindent
The appearance of a Fermi-Dirac factor (instead of the expected 
Planckian distribution) is known to occur in the response of a 
comoving Unruh-DeWitt detector (that is coupled to a massless 
scalar field) in odd-dimensional de Sitter spacetimes (see, for 
e.g., Refs.\cite{ooguri86,takagi85b}; for a recent discussion, 
see Ref.\cite{mty01}) and also in the case of a detector stationed 
at a constant radius in the spacetime of the $(2+1)$~dimensional 
Banados-Teitelboim-Zanelli (BTZ) black hole\cite{lo94}. 
Moreover, it has recently been pointed out that, not only comoving,
but even accelerated Unruh-DeWitt detectors in de Sitter spacetime 
as well as accelerated detectors (with proper acceleration beyond a 
certain critical value) in anti-de Sitter spacetime exhibit a thermal 
response\cite{dl97,jacobson98,dl98,dl99}. 
It will be interesting to investigate as to how the non-linearity
of the detector's coupling would affect the response of a static
detector around the BTZ black hole and also the response of comoving 
as well as accelerated detectors in de Sitter and anti-de Sitter 
spacetimes in different dimensions.
Furthermore, it has been shown that a similar ``inversion of 
statistics" occurs in the response of a monopole detector that 
is coupled linearly to the scalar density of a massless Dirac 
field in odd-dimensional flat and de Sitter 
spacetimes\cite{takagi86,terashima99,takagi85b,mty01}
and also around the BTZ black hole\cite{hsy95}, i.e. the response 
of the detector exhibits a Bose-Einstein factor instead of the 
Fermi-Dirac factor expected in such situations.
It will be worthwhile to examine as to how detectors coupled
non-linearly to the scalar density of spinor fields respond 
in odd-dimensional spacetimes. 
We plan to address these issues in some detail in a forthcoming 
publication\cite{sriramwp}.

\nonumsection{Acknowledgments}
\noindent
We would like to thank Don Page, Valeri Frolov and Jonathan 
Oppenheim for discussions, William Unruh for correspondence 
and discussions, Andrei Zelnikov for comments on an earlier
version of the manuscript and Supratim Sengupta for help with 
Maple.
We would also like to thank Christopher Stephens for correspondence 
and for providing a copy of his unpublished paper.
We also wish to thank the organizers of IGQR-I for their invitation
to speak at the Workshop. 
This work was supported by the Natural Sciences and Engineering 
Research Council of Canada. 

\nonumsection{References}

\end{document}